# Лавинный режим переключения перенапряженных высоковольтных *p-i-n*-диодов в проводящее состояние при импульсном освещении.

А. С. Кюрегян

Всероссийский Электротехнический институт им. В. И. Ленина, 111250, Москва, Россия
E-mail: ask@vei.ru

Построена простая аналитическая теория пикосекундного переключение высоковольтных перенапряженных *p-i-n*-фотодиодов в проводящее состояние при воздействии импульсного освещения. Получены и подтверждены путем численного моделирования соотношения между параметрами структуры, светового импульса, внешней цепи и основными характеристиками процесса – амплитудой импульса тока активной нагрузки, временем задержки и длительностью процесса коммутации. Показано, что необходимая для эффективной коммутации энергия пикосекундного импульса света может быть уменьшена на 6-7 порядков за счет интенсивного лавинного размножения электронов и дырок.

## Введение

Недавно в работе автора [1] был теоретически исследован процесс пикосекундного переключения высоковольтных обратносмещенных $p^+-n-n^+$-структур в проводящее состояние при воздействии короткого импульса света. Не вдаваясь в детали, основной результат работы [1] можно сформулировать так. Если длительность освещения $t_{ph}$ заметно меньше времени коммутации $t_k$, а энергия импульса света

$$W_{ph} > 2W_0 = 2\xi \frac{\hbar\omega E_0 d^2}{q u_s}, \qquad (1)$$

то

$$t_k \approx \tau \frac{W_0}{W_{ph}} = \frac{1}{n_0} \frac{\varepsilon E_0 S_d}{q u_s S_0}, \qquad (2)$$

где $n_0 = W_{ph}/\hbar\omega\xi S_0 d$ и $E_0$ - усредненные по толщине *n*-слоя $d$ начальная концентрация электронов и напряженность поля, несколько меньшая пробивного значения $E_b$, $\xi^{-1} = (1-R_{ph})(1-e^{-\kappa d})$, $\hbar\omega$ - энергия кванта, близкая к ширине запрещенной зоны, $R_{ph}$ и $\kappa$ - коэффициенты отражения и поглощения света, $u_s = (u_{sn}+u_{sp})$, $u_{sn,sp}$ - насыщенные дрейфовые скорости электронов и дырок, $\tau = R_L C_d$, $R_L$ - сопротивление нагрузки, $C_d = \varepsilon S_d/d$ - емкость диода, $S_d$ - площадь диода, $q$ - элементарный заряд, $\varepsilon$ - диэлектрическая проницаемость кремния. При использованных в [1] значениях параметров фотодиода $W_0 \approx 2.4$ мкДж и $\tau \approx 100$ пс, так что для эффективного управления необходимы импульсы света с энергией не менее 5 мкДж и длительностью не более 50 пс.

Такие параметры импульсов легко получить с помощью современных коммерческих неодимовых [2] и иттербиевых лазеров [3], энергия кванта которых почти идеально подходит для кремния. Однако их высокая стоимость, значительные габариты и низкий КПД являются совершенно неприемлемыми для многих практически важных применений. По этим показателям самыми подходящими источниками света являются импульсные полупроводниковые лазеры [4], но в настоящее время их мощность меньше необходимой на 5-6 порядков (см., например, [5]). В настоящей работе мы рассмотрим один из возможных вариантов разрешения этого противоречия. Именно, будет показано, что использование сильно перенапряженных (то есть смещенных в обратном направлении до напряжения $U_0 = (1.5 - 2.5) E_b d$) высоковольтных $p^+-i-n^+$-фотодиодов (которые мы будем называть лавинными фотодиодами - ЛФД) позволяет уменьшить необходимую для эффективной коммутации величину $W_{ph}$ на 6-7 порядков за счет очень интенсивного лавинного размножения электронов и дырок, порожденных пикосекундным импульсом света.

## Аналитическая теория

Рассмотрим процесс коммутации ЛФД, подключенного к источнику напряжения через последовательное сопротивление $R_L$. Будем считать, что собственная концентрация $n_i \ll \tau_g u_s / d^2 S_d$ ($\tau_g$ - генерационное время жизни). Это условие[1] обеспечивает полное отсутствие в истощенном $i$-слое электронов и дырок до начала освещения и возможность достичь значительного перенапряжения без пробоя диода за время $d/u_s$ порядка единиц наносекунд. Необходимая для этого техника давно отработана – см., например, [1]. После того, как напряжение $U$ на ЛФД достигнет величины $U_0 = (1.5 - 2.5) E_b d$ он освещается импульсом света и начинается процесс коммутации. Основное уравнение, которое его описывает, имеет вид [1]

$$C_d \frac{dU}{dt} = \frac{U_0 - U}{R_L} - S_o \bar{j}, \qquad (3)$$

где $S_0$ - освещаемая площадь ЛФД, $\bar{j}$ - усредненная по толщине $i$-слоя плотность тока электронов и дырок в освещенной области. Строго говоря, для вычисления функции $\bar{j}(t)$ нужно решить уравнение (3) вместе с уравнением Пуассона и уравнениями непрерывности для концентраций электронов $n(t,x)$ и дырок $p(t,x)$, учитывающих интенсивную ударную ионизацию. Сильная нелинейность такой задачи исключает возможность ее точного аналитического решения. Однако в ней содержится несколько потенциально малых параметров, что позволяет радикально упростить ситуацию. Далее будем считать, что

- длительность импульса света $t_{ph}$, много меньше времени задержки коммутации (см. далее);
- концентрация заряженных примесей в $i$-слое пренебрежимо мала;
- длина поглощения света $\kappa^{-1}$ в $i$-слое много меньше его толщины $d$;
- длительность всего процесса коммутации много меньше $d/u_s$.

В этом случае напряженность поля $E$ и концентрации носителей заряда $n = p$ не зависят от координаты почти во всем нелегированном $i$-слое [1]. Поэтому (3) можно заменить на

$$\frac{dE}{dt} = \frac{E_0 - E}{\tau} - \frac{S_o}{S_d} \frac{qnu}{\varepsilon}, \qquad (4)$$

и дополнить единственным уравнением

$$\frac{dn}{dt} = n\nu, \qquad (5)$$

где, $\nu = (\nu_n + \nu_p)$, $\nu_{n,p}$ - частоты ударной ионизации электронами и дырками, $u = (u_n + u_p)$, $u_{n,p}$ - дрейфовые скорости электронов и дырок. Первое начальное условие для этих уравнений

$$E(0) = E_0, \qquad (6)$$

---

[1] При типичных значениях параметров оно выполняется при температуре $T \leq 200$ K для кремния и $T \leq 660$ K для 4H-SiC.

означает, что за время освещения емкость диода не успевает перезарядиться током $qun_0S_0$, а второе имеет вид

$$n(0) = n_0. \tag{7}$$

При использовании для функций $u_{n,p}(E)$ и $v_{n,p}(E)$ даже самых простых аппроксимаций

$$u_{n,p} = u_{sn,sp}E(E+E_{sn,sp})^{-1}, \qquad v_{n,p} = \tilde{\alpha}_{n,p}u_{sn,sp}\exp(-\tilde{E}_{n,p}/E),$$

(здесь $E_{sn,sp}$, $u_{sn,sp}$, $\tilde{\alpha}_{n,p}$, $\tilde{E}_{n,p}$ - подгоночные параметры, значения которых для кремния приведены, например, в [8,8]) решение уравнений (4),(5) возможно только численными методами. Примеры таких решений приведены на **Рис. 1** (линии). Здесь и далее мы приводим результаты конкретных расчетов для кремниевого ЛФД при $T = 200$ K, $R_L = 50$ Ом, $\kappa = 5$ см$^{-1}$, $R_{ph} = 0.5$, $S_d = 9{,}44$ мм$^2$ и $S_0 = 0.5S_d$. Переменными являются параметры $U_0$, $d$ (или $U_b$) и $n_0$ (или $W_{ph}$).

Как видно, весь процесс коммутации разбивается на две качественно различных стадии. При $t < t_1$ напряжение на ЛФД практически не изменяется, поэтому в уравнении (5) $v(E) \approx v(E_0) \equiv v_0$, а в уравнении (4) $u(E) \approx u(E_0) \approx u_s$. С учетом этого получается решение

$$n(t) = n_0 \exp(v_0 t), \tag{8}$$

$$E(t) = E_0 - \frac{qS_0R_Lu_sn(t)}{d}\frac{1-\exp[-(v_0+1/\tau)t]}{1+\tau v_0}, \tag{9}$$

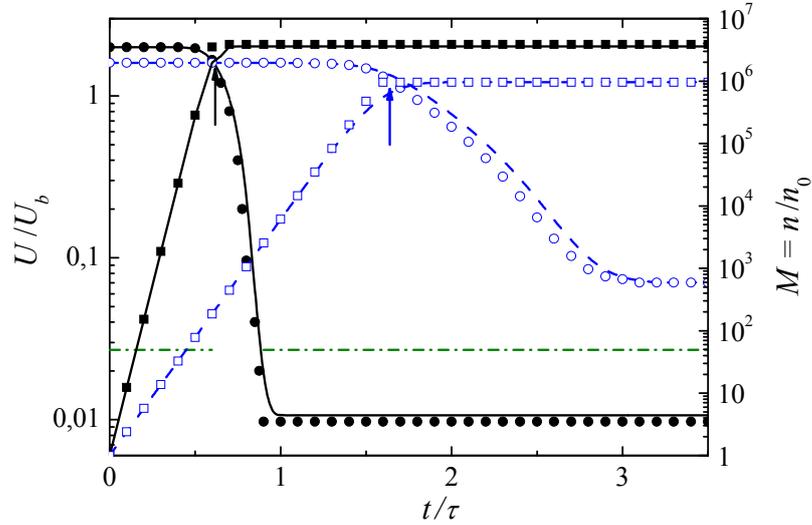

Рис. 1. Зависимости напряжения на ЛФД (кружки) и коэффициента размножения (квадраты) от времени, полученные путем точного (линии) и приближенного (символы) решения задачи (5)-(6) при $d = 250$ мкм, $U_b = 3{,}6$ кВ, $n_0 = 10^9$ см$^{-3}$, $\tau = 200$ пс и двух значениях $U_0 = 2U_b$ (сплошные линии и темные символы), $U_0 = 1.6U_b$ (штриховые линии и светлые символы). Стрелками указаны моменты времени $t_1$. Штрихпунктирной линией отмечено напряжение, при котором $E = E_{sn} \approx 0.4E_{sp}$.

которое остается справедливым до тех пор, пока[2]

$$\frac{\tilde{E}}{E} - \frac{\tilde{E}}{E_0} < \delta \approx 1$$

и скорость ударной ионизации заметно не уменьшится. При $t = t_1$ это неравенство превращается в равенство, откуда

$$E(t_1) \equiv E_1 = \frac{E_0 \tilde{E}}{\tilde{E} + E_0} \tag{10}$$

Комбинируя (8)-(10) можно показать, что

$$t_1 = \nu_0^{-1} \ln M_1, \tag{11}$$

а концентрация электронов (и дырок) при $t = t_1$ равна

$$n_0 M_1 \equiv n_1 = \frac{E_0^2}{\tilde{E} + E_0}\left(\frac{1}{\tau} + \nu_0\right)\frac{S_d}{S_0}\frac{\varepsilon}{q u_s}. \tag{12}$$

При выводе этих формул мы пренебрегали в (9) слагаемым $\exp\left[-(\nu_0 + 1/\tau)t_1\right]$, которое всегда меньше $M_1^{-1}$. В течение этой стадии ток $I_L = (U_0 - U)/R_L$ экспоненциально увеличивается со временем, но падение напряжения на нагрузке $I_L R_L$ остается много меньше $U \approx U_0$. Поэтому фактически время $t_1$ является временем задержки коммутации. Интересно отметить, что окончательные концентрации электронов и дырок $n_1 = n_0 M_1$ не зависят от $n_0$. Коэффициент лавинного размножения $M_1$ экспоненциально увеличивается, с ростом перенапряжения и может быть очень большим (см. **Рис. 2**).

При $t \approx t_1$ концентрация носителей заряда очень быстро увеличивается, а скорость ионизации – уменьшается. Поэтому практически сразу после завершения стадии задержки наступает вторая стадия, в течение которой $n(t) \approx M_1 n_0$, а напряженность поля еще больше $E_{sn,sp}$ и поэтому по-прежнему $u \approx u_s$. В этом случае решение уравнения (4) с «начальным» условием (10) имеет вид

$$E(\theta) = E_1\left[1 - \frac{E_0}{\tilde{E}}\tau\nu_0\left(1 - \exp\frac{t_1 - t}{\tau}\right)\right]. \tag{13}$$

Оно может быть использовано при всех $t > t_1$, если конечная напряженность поля

$$E(\infty) = E_1\left(1 - \frac{E_0}{\tilde{E}}\tau\nu_0\right) = E_0\frac{\tilde{E} - E_0\tau\nu_0}{\tilde{E} + E_0} \tag{14}$$

больше $E_{sn,sp}$, то есть при

$$\tau\nu_0 < \frac{\tilde{E}}{E_0}\left(1 - \frac{E_{sn,sp}}{E_0}\right) \approx \frac{\tilde{E}}{E_0}. \tag{15}$$

В сильном поле это неравенство не выполняется, условие $E > E_{sn,sp}$ нарушается при

$$t = t_s \approx t_1 - \tau\ln\left(1 - \tilde{E}/E_0\tau\nu_0\right),$$

---

[2] Излагаемая далее приближенная теория наилучшим образом согласуется с точным решением уравнений (4),(5) при варьировании $\delta$ в пределах (0,8-1). Эта неопределенность, очевидно, связана с неопределенностью выбора величины $\tilde{E}$ между $\tilde{E}_n$ и $\tilde{E}_p$. Конкретные расчеты мы выполняли при $\tilde{E} = \tilde{E}_n$ и $\delta = 0.9$, но в последующих формулах для их упрощения будем считать, что $\delta = 1$.

после чего напряженность поля быстро приближается к

$$E(\infty) = \frac{u_s}{\mu} \frac{1 + \tilde{E}/E_0}{1 + \tau \nu_0} < E_{sn,sp} \tag{16}$$

и вторая стадия завершается. При этом почти все напряжение источника (формирующей линии) перераспределяется с ЛФД на нагрузку, через которую протекает ток

$$I_M = \frac{U_0 - E(\infty)d}{R_L} \tag{17}$$

Если выполнено неравенство (15), то длительность этой стадии $t_2 \sim 2\tau$, а в противном, наиболее интересном случае

$$t_2 = t_s - t_1 \approx -\tau \ln\left(1 - \tilde{E}/E_0 \tau \nu_0\right). \tag{18}$$

При достаточно больших перенапряжениях, когда $E_0 \tau \nu_0 \gg \tilde{E}$,

$$t_2 \approx \frac{\tilde{E}}{E_0 \nu_0} \approx \frac{1}{n_1} \frac{\varepsilon E_0 S_d}{q u_s S_0} \tag{19}$$

и не зависит ни от параметров ЛФД, ни от сопротивления нагрузки, ни от $n_0$ (см. формулу (12)). Однако зависимость $t_2$ от $n_1 = n_0 M_1$ точно совпадает с (2). Это совпадение отражает тот факт, что собственно процесс коммутации (стадия 2) перенапряженным фотодиодом происходит точно так, как описано в [1], но начинается не сразу после окончания освещения, а через время задержки $t_1$, необходимое для увеличения начальной концентрации $n_0$ в $M_1$ раз вследствие интенсивной ударной ионизации. Пример зависимости $t_2$ и «инженерного» времени[3] коммутации $t_e$ приведен на **Рис. 2**.

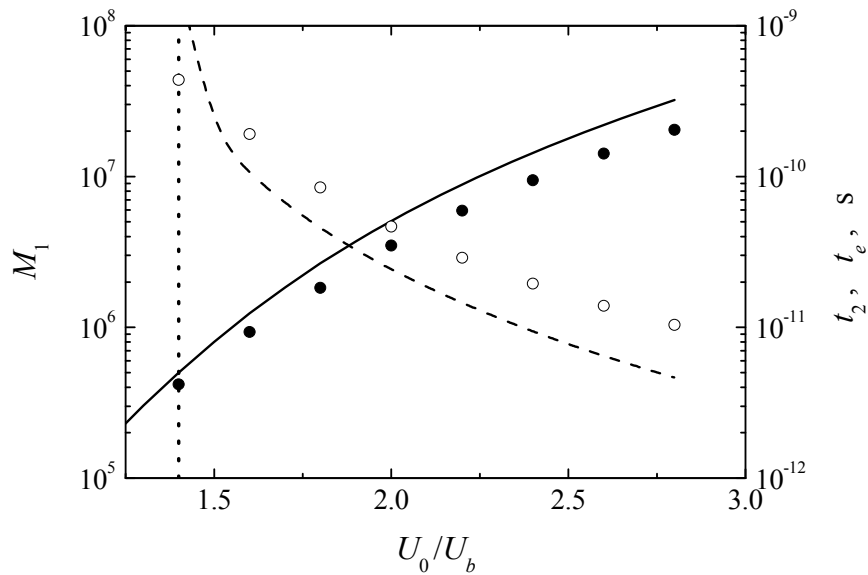

Рис. 2. Зависимость коэффициента размножения $M_1$ (сплошная линия и темные символы), длительности второй стадии коммутации $t_2$ (штриховая линия) и «инженерного» времени коммутации $t_e$ (светлые символы) от перенапряжения $U_0/U_b$ при $d = 250$ мкм, $n_0 = 10^9$ см$^{-3}$ или $W_{ph} = 1.5$ пДж. Символы – результаты точного решения уравнений (4) и (5), сплошная линия - расчет по формуле (12), штриховая линия - расчет по формуле (18). Вертикальная пунктирная линия отмечает значение $U_0/U_b$, при котором нарушается условие (15).

---

[3] Это время, за которое ток нагрузки увеличивается от $0.1 I_M$ до $0.9 I_M$

## Обсуждение результатов

Сравнение результатов точного и приближенного решения уравнений (4),(5), представленных на **Рис. 1** и **Рис. 2** показывает, что формулы (12)-(18) неплохо описывают процесс коммутации. В свою очередь результаты точного решения уравнений (4),(5) прекрасно согласуются с результатами численного моделирования, проведенного с помощью программы «Исследование» [9] (см. **Рис. 3**), вплоть до фактического завершения процесса коммутации. Однако приближенный расчет $E(\infty)$ дает сильно заниженное значение остаточного напряжения. Причина этого состоит в том, что в уравнениях (4),(5) не учтено (в силу неравенства $t \ll d/u_s$ - см. выше) обеднение приграничных областей $i$-слоя носителями заряда, дрейфующими внутрь $i$-слоя [1]. Толщина этих областей мала, но плотность объемного заряда в них (порядка $qn_1$) велика, так что сумма «катодного» и «анодного» напряжений может превысить $E(\infty)d$ даже при $t \sim t_1 \ll d/u_s$. Дополнительный (но обычно незначительный [1]) вклад в остаточное напряжение дает продольное сопротивление растекания $p^+$-слоя в окне фотодиода.

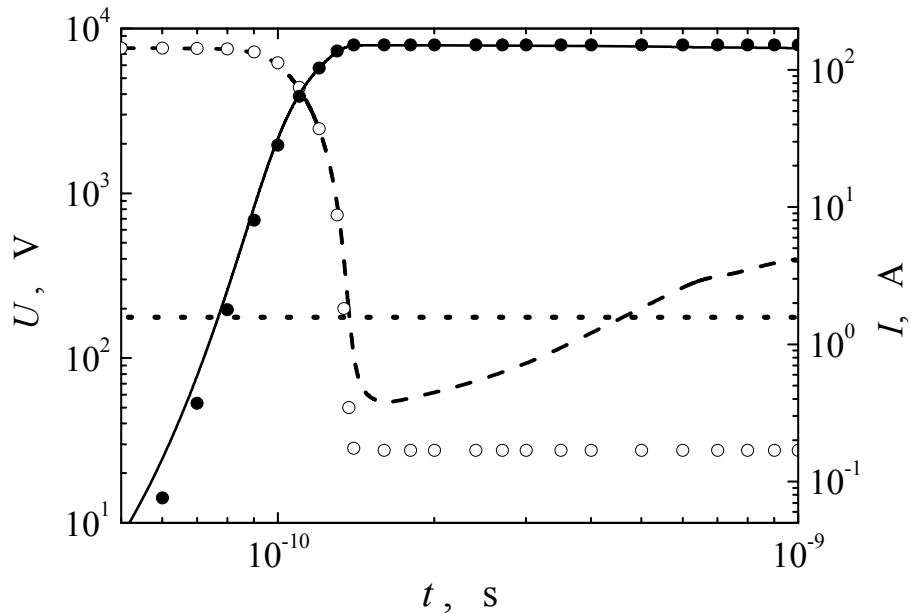

Рис. 3. Зависимости тока через нагрузку (сплошная линия и темные символы) и напряжения на ЛФД (штриховая линия и светлые символы) от времени, полученные путем решения задачи (5)-(6) (символы) и точного численного моделирования (линии) для значений $d = 250$ мкм, $U_b = 3{,}8$ кВ, $U_0 = 2U_b$, $n_0 = 10^9$ см$^{-3}$. Горизонтальной пунктирной линией отмечено напряжение $E_{sn}d$.

Еще одно ограничение применимости результатов состоит в том, что все они были получены в континуальном приближении, которое, нарушается при достаточно малых значениях $n_0$. Действительно, к моменту $t$ объем каждой лавины, порожденной одним электроном или дыркой, достигает величины $4\pi D u_s t^2$ [10-12], где $D$ - коэффициент поперечной диффузии. Эти лавины перекрываются при

$$t = t_{ov} = \left(4\pi n_0 D u_s\right)^{-1/2}. \tag{20}$$

Если $t_{ov} < t_1$, то континуальное приближение заведомо применимо. Это условие нарушается при больших перенапряжениях, так как с ростом начальной напряженности поля $E_0$ время задержки уменьшается (см. **Рис. 4**). Однако если среднее расстояние между лавина-ми $(2n_0)^{-1/3} \ll d$, но время лавинно-стримерного перехода $t_a > t_1$, то континуальное приближение по-прежнему применимо, хотя и с меньшей точностью. Зависимость $t_a(E_0)$ для Si при 300 К, полученная в работе [8], хорошо описывается функцией

$$t_a = t_{a0} \exp(E_a/E), \qquad (21)$$

где $t_{a0} = 1.0$ пс, $E_a = 1.37$ МВ/см. При 200 К параметр $E_a$ должен быть уменьшен примерно на 7%. Скорректированная таким образом зависимость $t_a(E_0)$ изображена на Рис. 4. Как видно, лавины могут превратиться в стример только при очень больших перенапряжениях ($U_0 \geq 2U_b$) и малых начальных концентрациях ($n_0 \leq 10^7$ см$^{-3}$). Если же $n_0 \geq 10^9$ см$^{-3}$, то все условия применимости теории оказываются выполненными при $U_0 \leq 3U_b$.

Таким образом, можно ожидать, что использование лавинного режима коммутации высоковольтных фотодиодов позволит применить для управления импульсные полупроводниковые лазеры. В качестве примера, иллюстрирующего возможности этого режима, приведем оценку для прибора с $d = 250$ мкм, $U_b = 3570$ В и $U_0 = 2U_b$. Под действием импульса света с длительностью менее 20 пс, энергией 1,5 пДж и коэффициентом поглощения 5 см$^{-1}$ такой фотодиод должен сформировать на активной 50-омной нагрузке импульс с мощностью около 1 МВт, задержкой 100 пс и фронтом 46 пс.

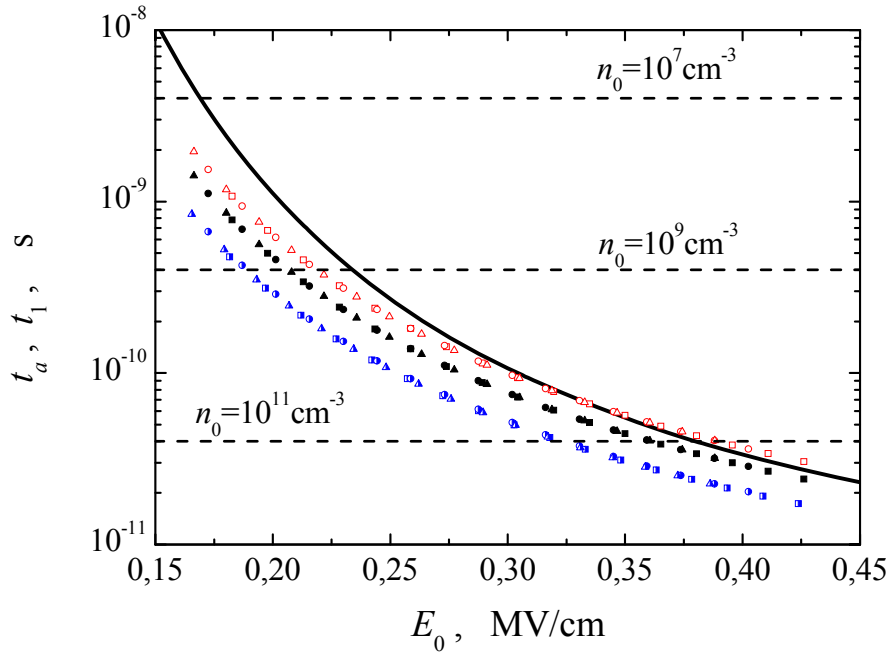

Рис. 4. Зависимости времени лавинно-стримерного перехода $t_a$ (линия, по данным [8], $T = 300$ К) и времени задержки процесса коммутации $t_1$ (символы, расчет по формуле (11)) от напряженности поля $E_0$ при $n_0 = 10^7$ см$^{-3}$ (светлые символы), $10^9$ см$^{-3}$ (темные символы), $10^{11}$ см$^{-3}$ (светло-темные символы) и $d = 150$ мкм (квадраты, $E_b = 151$ кВ/см), 250 мкм (кружки, $E_b = 143$ кВ/см), 350 мкм (треугольники, $E_b = 138$ кВ/см). Штриховыми линиями отмечено время перекрытия лавин $t_{ov}$, рассчитанное по формуле (20) при $D = 10$ см$^2$/с и различных $n_0$.